# Drift velocity saturation and large current density in an intrinsic three-dimensional Dirac semimetal cadmium arsenide

S. S. Kubakaddi[a]
Department of Physics, K. L. E. Technological University, Hubballi-580031, Karnataka, India
(November 20, 2019)

**Abstract**

Transport of electrons at high electric fields is investigated in an intrinsic three-dimensional Dirac semimetal cadmium arsenide, considering the scattering of electrons from acoustic and optical phonons. Screening and hot phonon effect are taken in to account. Expressions for the hot electron mobility $\mu$ and power loss $P$ are obtained as a function of electron temperature $T_e$. The dependence of drift velocity $v_d$ on electric field $E$ and electron density $n_e$ has been studied. Hot phonon effect is found to set in the saturation of $v_d$ at relatively low $E$ and significantly degrades its magnitude. The drift velocity is found to saturate at a value $v_{ds} \sim 10^7$ cm/s and it is weakly dependent on $n_e$. A large saturation current density $\sim 10^6$ A/cm$^2$ is predicted.

## 1. Introduction

Three-dimensional (3D) Dirac semimetal (3DDS) cadmium arsenide, a 3D analog of graphene, is one of the most emergent class of materials in current condensed matter physics (for review see Refs. [1, 2]). Interest in the study of electronic properties of 3DDS cadmium arsenide (Cd$_3$As$_2$) was initiated by its theoretical prediction [3], and experimental realization [4-8]. The linear band dispersion with gapless feature has led massless Dirac fermions in this material to exhibit many unusual transport phenomena such as quantum oscillations [9-10], ultrahigh mobility [8,10,11] and giant magnetoresistance [8,11,12]. Moreover, recently quantum Hall effect has been observed in films of Dirac semimetal Cd$_3$As$_2$ [13,14].

Very large mobilities > ~ 10$^7$ cm$^2$/V-s have been reported in 3DDS Cd$_3$As$_2$ at low- temperature ~ 5 K [6-8, 10]. Resistivity measurements as a function of temperature $T$ have shown metallic behavior with the residual resistivity as low as 11.6 n$\Omega$ cm at $T <$ 6 K [10]. This ultra-large (small) mobility (resistivities) are attributed to the suppressed back scattering of high-velocity massless Dirac fermions. In a sample with impurities, the residual mobility has been found to be ~ 2×10$^5$ cm$^2$/V-s at $T \sim$ 2.6 K and it decreases with increasing temperature, reaching finally a value of ~ 2×10$^4$ cm$^2$/V-s at room temperature [15]. Some recent experiments have also shown room temperature mobility > ~ 10$^4$ cm$^2$/Vs [16, 17]. In addition, in 3DDS Cd$_3$As$_2$, the electron density $n_e$ is ultra-large (~10$^{18}$ – 10$^{20}$ cm$^{-3}$) [7, 15, 18, 19], exceeding the densities in traditional semiconductors. The behavior of temperature dependence of the resistivity/mobility was inferred to be due to the impurity scattering at low-temperature and phonon scattering at higher temperature.

There have been a few theoretical investigations of the transport properties in 3DDS Cd$_3$As$_2$ considering the scattering by short-range disorder and charged impurities using the Boltzmann transport equation [20, 21]. Very recently, we have given a theory of the phonon-limited mobility of high density Dirac fermions in 3DDS Cd$_3$As$_2$ considering the electron scattering by acoustic and optical phonons in the quasi-elastic approximation [22]. Experimental results of the mobility have been quantitatively explained by applying this theory.

The above mentioned transport studies have been made in the low electric field. In the high electric field, hot electron cooling as a function of electron temperature has been theoretically investigated [23-25]. However, there have been no reports, either experimentally or theoretically, giving the electron drift velocity $v_d$ dependence of electric field $E$ in the high field region where $v_d$ tends to saturate. In the Drude model, the current density is given by $j = n_e e\, v_d$. Realization and control of saturation drift velocity $v_{ds}$ /current density $j_s$ is an important measure for the device applications of field effect transistors (FETs) at high electric field. FETs for analog and radiofrequency (RF) applications ideally operate in the saturation limit. Because of the excellent carrier mobility at room temperature and large $n_e$, the 3DDS Cd$_3$As$_2$ is expected to give rise to a large saturation current density. Thus, Cd$_3$As$_2$ can be considered as a potential candidate in high efficiency RF analog devices.

Very recently, the two-dimensional Dirac fermions have been realized in films of Cd$_3$As$_2$. [13, 26, 27]. FETs using Cd$_3$As$_2$ film of thickness 30 nm as the channel material are demonstrated [27]. These FETs have shown extremely high current densities and low contact resistances and are very promising for future high-speed electronics and RF applications.

In view of the above observations, an understanding of the velocity-field characteristics in 3DDS $Cd_3As_2$ is very essential. These characteristics have been extensively studied in bulk semiconductors [28-30], low-dimensional semiconductors [31] and graphene [32-37]. In the present work we have conducted a study of drift velocity and current density behaviour as a function of electric field and electron density in 3DDS $Cd_3As_2$. The hot electron mobility and energy balance equations are employed to explore this behaviour. Since the nearly intrinsic samples have been realized for $T > \sim 5$ K, [6-8,10], we have considered phonons as the only scattering channel.

The theory of hot electron intrinsic mobility is developed for the first time in 3DDS by including the hot phonon effect and screening. It is presented, along with the hot electron power loss equations, in section II. Using these equations, the velocity-field curves are obtained by numerical solution. The results and discussion are presented in section III. Finally, our findings are summarised in section IV.

## 2. Analytical model for high field transport in three-dimensional Dirac semimetal

We consider Dirac fermions in a disorder free 3DDS $Cd_3As_2$ with a large electron density so that the Fermi energy $E_F$ is well above the Dirac points. The electron energy dispersion is linear, i.e., $E_k = \hbar v_F k$, and the density of states is $D(E_k) = gE_k^2/[2\pi^2(\hbar v_F)^3]$, where $v_F$ is the Fermi velocity, $k$ is the 3D wave vector, and $g = g_s g_v$, with $g_s$ ($g_v$) denoting the spin (valley) degeneracy of the electron. We assume, that the electronic dispersion is isotropic [4, 21- 23] although it has been found by some authors to be anisotropic [3, 7,8]. In an applied electric field $E$ electrons gain energy and momentum, and in the steady state they establish their electron temperature $T_e$ (> $T$, the lattice temperature) and drift velocity $v_d$, by losing extra energy and momentum to the lattice (phonons). The electrons are assumed to obey the hot-electron Fermi-Dirac (F-D) distribution $f_o(E_k) = \{\exp[(E_k - E_F)/k_B T_e] +1\}^{-1}$. The three parameters $T_e$, $v_d$ and the hot electron mobility $\mu = v_d / E$ give a full description of the 3D Dirac electrons in non-equilibrium.

We assume that electrons interact with the intrinsic acoustic phonons (ap) and optical phonons (op). Since we consider a $Cd_3As_2$ with large $n_e \sim (10^{18} - 10^{20}$ cm$^{-3})$, the electron scattering by both ap and op is assumed to be quasi-elastic, noting that the optical phonon energy $\sim 25$ meV [38] is considerably smaller than the $E_F$. Hence, we can obtain the phonon-limited hot electron intrinsic mobility $\mu$ using the semi-classical Boltzmann transport equation solved in the relaxation time approximation. The mobility, thus obtained, will be a function of $T_e$. In the steady state, the energy balance equation is given by $eEv_d = P$, where $eEv_d$ is the power gained by electron from the field $E$ and $P = <dE_k/dt>_{el-ph}$ is the power loss per electron to the lattice by electron-phonon (el-ph) interaction. $P$ can be calculated by the standard technique [25, 28] and it will be a function of $T_e$. Combining the equation for drift velocity $v_d = \mu E$ and $eEv_d = P$ an expression relating $T_e$ to $E$ can be obtained. Hence, $v_d$ vs $E$ curves are deduced.

### 2.1 Phonon-limited hot electron mobility $\mu$

From the Boltzmann transport equation technique, using the relaxation time approximation, an expression for the mobility is given by [22]

$$\mu_i = \sigma_i / n_e e, \text{ with electrical conductivity } \sigma_i = e^2 K_{0i}, \quad (1)$$

where,

$$K_{0i} = \frac{v_F^2}{3}\int dE_k D(E_k)\tau_i(E_k)\left(-\frac{\partial f_o(E_k)}{\partial E_k}\right), \quad (1a)$$

and $i$ = ap and op. For $E_F >> k_B T_e$, expression for the mobility takes the simple form $\mu_i = [ev_F^2 D(E_F)\tau_i(E_F)]/3n_e$.

Considering the electron interaction with the intrinsic phonons of energy $\hbar\omega_q$ and wave vector $\mathbf{q}$, the energy-dependent hot electron momentum relaxation time $\tau(E_k)$ for the scattering due to phonons, following Ref. [22], is shown to be

$$\frac{1}{\tau(E_k)} = \left(\frac{V}{2\pi\hbar(\hbar v_F)^3}\right)[1-f_o(E_k)]^{-1}\int_0^\pi d\theta(1-\cos\theta)]\sin\theta$$
$$\times \frac{|C(q)|^2}{\varepsilon^2(q)}\{N_q(E_k+\hbar\omega_q)^2[1-f_o(E_k+\hbar\omega_q)]+$$
$$(N_q+1)(E_k-\hbar\omega_q)^2[1-f_o(E_k-\hbar\omega_q)]\theta(x)\}, \quad (2)$$

where $V$ is the volume of the crystal, $\theta$ is the angle between the initial state $\mathbf{k}$ and final state $\mathbf{k}'$, $|C(q)|^2$ is the electron-phonon matrix element, $\varepsilon(q)$ is the screening function, $N_q$ is the phonon distribution function, and $\theta(x)$ is the step function with $x = (E_k - \hbar\omega_q)$. We take the temperature independent screening function $\varepsilon(q) = [1+(q_{TF}/q)^2]$ in the Thomas-Fermi approximation, where $q_{TF} = [4\pi e^2 D(E_F)/\varepsilon_s]^{1/2}$ is the Thomas-Fermi wave vector [21]. This is valid for $T_e << T_F$, where $T_F$ is the Fermi temperature. In the following, we obtain the hot electron momentum relaxation time due to the scattering by acoustic and optical phonons.

### 2.1.1 *Hot electron momentum relaxation time due to acoustic phonon scattering*

The electron scattering by acoustic phonons is taken to be via deformation potential coupling. The corresponding interaction matrix element is given by $|C(q)|^2 = [(D^2\hbar\omega_\mathbf{q})/(2\rho_m V v_s^2)](1+cos\theta)/2$ [21,23], where $D$ is the acoustic deformation potential constant, $\rho_m$ is the mass density, $\omega_\mathbf{q} = v_s q$, and $v_s$ is the velocity of acoustic phonon. Assuming the acoustic phonons to be in thermal equilibrium with the lattice, then $N_\mathbf{q}$ is given by the Bose distribution $N_\mathbf{q}(T) = \{\exp[(\hbar\omega_\mathbf{q})/k_B T]-1\}^{-1}$ at lattice temperature $T$. Using the quasi-elastic approximation, the relaxation time $\tau_{ap}(E_\mathbf{k})$ for the acoustic phonon scattering is given by

$$\frac{1}{\tau_{ap}(E_\mathbf{k})} = \frac{D^2 v_F}{8\pi\rho_m v_s^2(\hbar v_s)^4 E_\mathbf{k}^2(k_B T_e)} \int_0^{2\hbar v_s k} d(\hbar\omega_\mathbf{q}) \frac{(\hbar\omega_\mathbf{q})^5}{\varepsilon^2(q)}$$

$$\times \left[1 - \left(\frac{\hbar\omega_\mathbf{q}}{E_\mathbf{k}}\right)^2 \left(\frac{v_F}{2v_s}\right)^2\right] [N_\mathbf{q}(T)+1] N_\mathbf{q}(T_e)$$

$$\times \left\{\exp\left[\left(\frac{\hbar\omega_\mathbf{q}}{k_B}\right)\left(\frac{1}{T_e} - \frac{1}{T}\right)\right] + 1\right\}. \quad (3)$$

We obtain the analytical simple forms of $\tau_{ap}(E_\mathbf{k})$ in special cases of very low temperature i.e. Bloch-Grüneisen (BG) regime and high temperature i.e. equipartition (EP) regime.

(i) *Very low temperature*: In the BG regime, $q \to 0$ as $T \to 0$, $\hbar\omega_\mathbf{q} \approx k_B T$, and $\hbar\omega_\mathbf{q} \ll k_B T_e$. We set $k=k_F$ (the Fermi wave vector), $E_\mathbf{k}=E_F$ and $\varepsilon(q)\approx (q_{TF}/q)^2$. Then, the momentum relaxation time in the BG regime is given by

$$\frac{1}{\tau_{ap-BG}(E_F)} = \frac{9! D^2(hv_F)}{8\pi\rho_m v_s(\hbar v_s)^5 E_F^2 (\hbar v_s q_{TF})^4}$$

$$\times \left[\frac{(k_B T_e)^{10} + (k_B T)^{10}}{k_B T_e}\right] \quad (4a)$$

with screening, and

$$\frac{1}{\tau_{ap-BG}(E_F)} = \frac{15 D^2(hv_F)}{\pi\rho_m v_s(\hbar v_s)^5 E_F^2} \left[\frac{(k_B T_e)^6 + (k_B T)^6}{k_B T_e}\right] \quad (4b)$$

without screening.

Thus, in 3DDS for $T_e \gg T$, we find that

$$\tau_{ap-BG}(E_F) \sim T_e^{-9} \text{ and } T_e^{-5}, \quad (4c)$$

respectively, for the screened and unscreened el-ph interaction. This $T_e$ dependence is same as in the conventional degenerate 3D semiconductor [39] and it is manifestation of the 3D nature of acoustic phonons, noting that screening is taken to be independent of temperature. Correspondingly, the acoustic phonon limited BG regime hot electron mobility gives the following $T_e$ dependence

$$\mu_{ap-BG} \sim T_e^{-9} (T_e^{-5}), \text{ with (without) screening.} \quad (4d)$$

It is to be noted that for the low field case, $T_e = T$, and Eqs.(4a)–(4d) reduce to those in Ref.[22]. We also find that $n_e$ dependence of hot electron $\mu_{ap-BG}$ is given by

$$\mu_{ap-BG} \sim n_e^{5/3} (n_e^{1/3}), \text{ with (without) screening.} \quad (4e)$$

It is the same as found for low field case in Ref. [22].
(ii) *High temperature*: In the EP regime, $\hbar\omega_\mathbf{q} \ll k_B T, k_B T_e$, the equation (3) simplifies to

$$\frac{1}{\tau_{ap-EP}(E_\mathbf{k})} = \frac{D^2 v_F(k_B T)}{8\pi\rho_m v_s^2(\hbar v_s)^4 E_\mathbf{k}^2} \int_0^{2\hbar v_s k} d(\hbar\omega_\mathbf{q}) \frac{(\hbar\omega_\mathbf{q})^5}{\varepsilon^2(q)}$$

$$\times \left[1 - \left(\frac{\hbar\omega_\mathbf{q}}{E_\mathbf{k}}\right)^2 \left(\frac{v_F}{2v_s}\right)^2\right] \left[2 + \frac{\hbar\omega_\mathbf{q}}{k_B}\left(\frac{1}{T_e} - \frac{1}{T}\right)\right]. \quad (5a)$$

For unscreened el-ap coupling, it may be approximated to give a simple form

$$\frac{1}{\tau_{ap-EP}(E_\mathbf{k})} = \frac{D^2 v_F(k_B T) E_\mathbf{k}^2}{3\pi\rho_m v_s^2(\hbar v_s)^4}, \quad (5b)$$

which is independent of $T_e$. Interestingly, in the EP regime, for $E_\mathbf{k}=E_F$ the $\tau_{ap-EP}(E_F)$ and the corresponding mobility $\mu_{ap-EP}$ are the same as in the low field case [22]. Consequently, $T$ and $n_e$ dependence of $\mu_{ap-EP}$ are also same as found for zero field case i.e.

$$\mu_{ap-EP} \sim T^{-1} \text{ and } n_e^{-1}. \quad (5c)$$

### 2.1.2 *Hot electron momentum relaxation time due to optical phonon scattering*

We consider the electron-optical phonon interaction via Fröhlich interaction and the corresponding matrix element is given by $|C(q)|^2 = (C_0/q^2)(1+cos\theta)/2$, where $C_0 = (2\pi e^2 \hbar\omega_0 \varepsilon')/V$, $\hbar\omega_\mathbf{q} = \hbar\omega_0$ is the optical phonon energy, $\varepsilon' = (\varepsilon_\infty^{-1} - \varepsilon_s^{-1})$, and $\varepsilon_\infty (\varepsilon_s)$ is the high-frequency (static) dielectric constant. In high electric field, the optical phonon distribution will deviate from its thermal equilibrium value $N_\mathbf{q}(T)$ and it is given by the hot phonon distribution function $N_{\mathbf{q}hp}$ [25]. Assuming the scattering to be quasi-elastic, the momentum relaxation time due to optical phonons, taking account of the hot phonon effect and screening, is found to be

$$\frac{1}{\tau_{op}(E_\mathbf{k})} = \frac{e^2 \hbar \omega_0 \varepsilon'}{2\hbar^2 v_F} \left(\frac{\hbar \omega_0}{k_B T_e}\right) \int_0^\pi d\theta \sin\theta \left(\frac{1+\cos\theta}{2}\right) \times$$

$$\times \left\{ \begin{array}{l} \frac{N_{\mathbf{q}_+ hp}[N_\mathbf{q}(T_e)+1](1+(\hbar\omega_0/E_\mathbf{k}))}{\varepsilon^2(q_+)} \\ + \frac{[N_{\mathbf{q}_- hp}+1]N_\mathbf{q}(T_e)[1-(\hbar\omega_0/E_\mathbf{k})]\theta(x)}{\varepsilon^2(q_-)} \end{array} \right\}, \quad (6)$$

where

$$q_+^2 = (1/\hbar v_F)^2 [2E_\mathbf{k}^2 + 2E_\mathbf{k}\hbar\omega_o + (\hbar\omega_o)^2 - 2E_\mathbf{k}(E_\mathbf{k}+\hbar\omega_o)\cos\theta] \quad (6a)$$

and

$$q_-^2 = (1/\hbar v_F)^2 [2E_\mathbf{k}^2 - 2E_\mathbf{k}\hbar\omega_o + (\hbar\omega_o)^2 - 2E_\mathbf{k}(E_\mathbf{k}-\hbar\omega_o)\cos\theta], \quad (6b)$$

and $N_{\mathbf{q}+hp}$ and $N_{\mathbf{q}-hp}$ are the hot phonon distribution functions with $q=q_+$ and $q=q_-$, respectively.

The hot electron mobility due to ap and op scattering can be obtained by using the respective relaxation times in Eq.(1). The resultant phonon-limited hot electron mobility is given by $\mu = [(1/\mu_{ap}) + (1/\mu_{op})]^{-1}$.

From the equation $v_d = \mu E$, the momentum balance equation can be obtained as $eE = (ev_d/\mu) = Q$, where $eE$ is the force on an electron due to electric field $E$. The substitution of phonon-limited hot electron mobility $\mu$, on r h s, gives the momentum loss rate $Q$ due to phonons. Conventionally, $Q$ is obtained by finding the average momentum loss rate $<d\hbar\mathbf{k}/dt>$ to the lattice using the displaced hot electron F-D distribution [28,40,41].

## 2.2 Hot electron power loss $P$

The hot electron power loss due to the intrinsic acoustic and optical phonons has been investigated by us in detail [24,25]. For completeness sake their final results are given here. The power loss due to acoustic phonons, with screened el-ap interaction, is given by

$$P_{ap} = -\frac{gD^2}{8\pi^3 \rho \hbar^7 n_e v_F^4 v_s^4} \int_0^\infty dE_\mathbf{k} \int_0^{(\hbar\omega_q)_{max}} d(\hbar\omega_\mathbf{q})(\hbar\omega_\mathbf{q})^3 \frac{(E_\mathbf{k}+\hbar\omega_\mathbf{q})^2}{\varepsilon^2(q)}$$
$$\times |F(E_\mathbf{k},E_\mathbf{q})|^2 [N_\mathbf{q}(T_e)-N_\mathbf{q}(T)][f_0(E_\mathbf{k})-f_0(E_\mathbf{k}+\hbar\omega_\mathbf{q})]. \quad (7)$$

Taking account of the hot phonon effect and screening of el-op interaction, the power loss due to optical phonons is shown to be

$$P_{op} = \frac{g(\hbar\omega_0)^2 e^2 \varepsilon'}{2\pi^2 \hbar n_e (\hbar v_F)^4} \int_0^\infty dE_\mathbf{k} E_\mathbf{k} \int_0^{E_{qu}} dE_\mathbf{q} \frac{|F(E_\mathbf{k},E_\mathbf{q})|^2}{E_\mathbf{q} \varepsilon^2(q)}$$
$$\times [(N_{\mathbf{q}hp}+1)e^{-(\hbar\omega_0/k_B T_e)} - N_{\mathbf{q}hp}]$$
$$\times f(E_\mathbf{k})[1-f(E_\mathbf{k}+\hbar\omega_0)]|E_\mathbf{k}+\hbar\omega_0|, \quad (8)$$

where

$$|F(E_\mathbf{k},E_\mathbf{q})|^2 = (1/2)\{1+[(E_\mathbf{k}^2 - E_\mathbf{q}^2)+(E_\mathbf{k}+\hbar\omega_0)^2] \times [2E_\mathbf{k}(E_\mathbf{k}+\hbar\omega_0)]^{-1}\} \quad (8a)$$

and $E_{qu} = (2E_\mathbf{k}+\hbar\omega_0)/2$. The equation (8) is obtained from our work [25] by combining Eqs.(16) and (6) and substituting for $|g(q)|^2 = [(2\pi e^2 \hbar\omega_0 \varepsilon')/Vq^2]$. The total power loss is given by $P = P_{ap} + P_{op}$. The energy balance equation is obtained as $eEv_d = P$.

It should be noted that the method adopted here to obtain $v_d$ as a function of $E$ is analytical, unlike other numerical methods [29, 32, 36, 37].

## 3. Results and Discussion

The above expressions for the hot electron mobility $\mu$ (hence for the drift velocity $v_d$) and the energy balance are the transcendental equations. We numerically solve these coupled equations for a chosen $T_e$ to obtain velocity - field curves in 3DDS Cd$_3$As$_2$. The material parameters used are: $v_F = 1\times10^8$ cm s$^{-1}$ [21- 23, 42], $v_s = 2.3\times10^5$ cm s$^{-1}$, $\rho_m = 7.0$ g/cm$^3$, $\varepsilon_\infty = 12$, $\varepsilon_s = 36$, and $g = 4$. A reasonable value of $D = 20$ eV [22, 43] and a typical value of $\hbar\omega_0 = 25$ meV [22, 38] are chosen. Throughout the discussion $n_0 = 1\times10^{18}$ cm$^{-3}$ is used. The reasonable values of hot phonon relaxation time $\tau_p = 1$ and 5 ps are used in the demonstration. These values of $\tau_p$ are of the order that have been experimentally shown [44] and used in theoretical calculations [37, 45] in graphene. Also, in III-V semiconductors $\tau_p$ is of the order of a few ps [40].

In order to analyze the $v_d$ vs $E$ characteristics, we need to know the $T_e$ dependence on $E$, because $P$ and $\mu$ are determined by $T_e$. A representative $T_e$ vs $E$ behavior is depicted in figure1a for $n_e= n_0$ and $5n_0$, $T= 4.2$ K and $\tau_p = 0$ ps. It has been found that $T_e$ deviates from $T$ and sets increasing rapidly at about $E = 0.03$ (0.1) V/cm for $n_e= n_0$ ($5n_0$). For $E > \sim 0.2$ and 0.6 V/cm, respectively, for $n_e= n_0$ and $5n_0$, the rate of increase of $T_e$ slows down. This is the region where electron-optical phonon scattering plays the dominant role as an energy dissipation channel. Similar observation is made in bulk InSb semiconductor whose optical phonon energy is 24.4 meV [46], closer to the one considered in the present work. In addition, $T_e$ is found to have a strong (weak) dependence on $n_e$ at low (large) $E$. The strong $n_e$ dependence at low field may be attributed to the $n_e$ dependence of $\mu_{ap}$ and $P_{ap}$. In figure 1b, $T_e$ vs $E$ is shown for $n_e = n_0$ for different $\tau_p$. In the low field region, where ap scattering is dominant, $T_e$ is independent of $\tau_p$. The hot phonon effect is found to enhance $T_e$ in the high field region. This is because, the number of hot phonons increases with increasing $T_e$. This increased number of hot phonons can be described by the Bose distribution $N_q$ ($T_{ph}$), with effective hot phonon temperature $T_{ph}$ being $T < T_{ph} < T_e$. The difference $T_e - T_{ph}$ decreases with increasing $\tau_p$, and reduces the electron energy loss. Consequently, $T_e$ is enhanced for larger $\tau_p$.

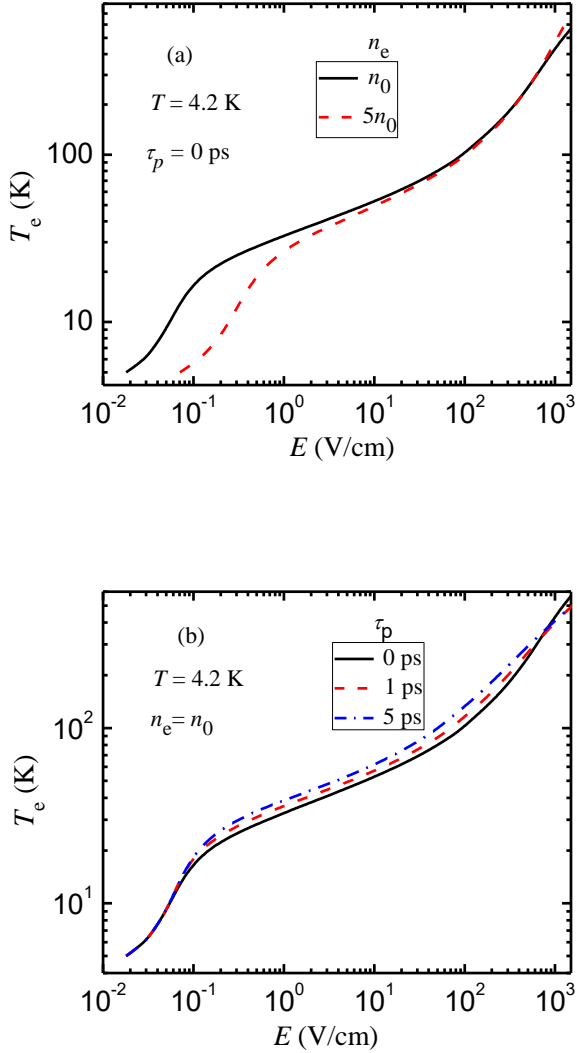

**Figure 1.** The electron temperature $T_e$ as a function of electric field $E$, at $T= 4.2$ K. (a) For electron densities $n_e = n_0$ and $5n_0$ and $\tau_p = 0$ ps and (b) for electron density $n_e = n_0$ and $\tau_p = 0$, 1 and 5 ps.

In figure 2, the drift velocity $v_d$ is plotted as a function of $E$, at $T = 4.2$ K, for $n_e = n_0$ and $\tau_p = 0$, 1 and 5 ps. The behavior is, as conventionally found, linear at very low field and becomes sub-linear for higher field. Finally, for further increase of $E$, $v_d$ saturates or tends to saturate. This behavior is found to be the same for all $\tau_p$. The saturation/ near saturation drift velocity has been found to be of the order of $10^7$ cm/s. It is nearly the same as found in graphene [32, 34, 35] and in the 3D conventional III-V semiconductors [29, 40]. The nonlinear behavior is generally attributed to the enhanced el-ph scattering with the increasing field. We may also explain the high field behavior using the energy and momentum balance equations. These equations imply that the drift velocity is given by $P/Q$. In order to understand the saturation / near saturation of $v_d$, we have to know the dependence of $P$ and $Q$ on $T_e$. At higher $T_e$ (i.e. at strong $E$), $P$ has been found to increase rapidly with $T_e$ [25], and hence with $E$. This is due to the enhanced scattering by optical phonons at high fields. Similar behavior is expected with Q. This rapid increase of momentum and energy loss rates results in saturation /near saturation of $v_d$.

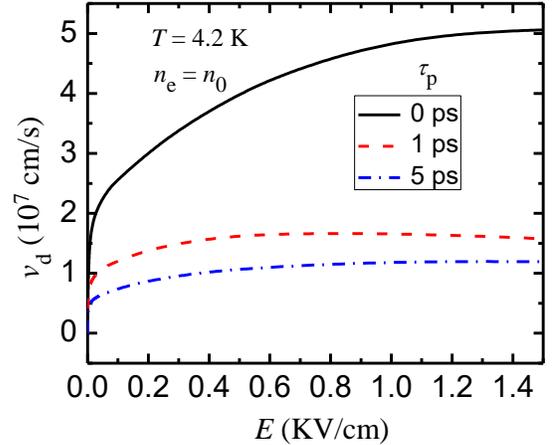

**Figure 2**. The electron drift velocity $v_d$ as a function of electric field $E$, at $T= 4.2$ K, for electron density $n_e = n_0$ and phonon relaxation time $\tau_p = 0{,}1$ and 5ps.

The hot phonon effect can be captured, in figure 2, by comparing the $v_d$ vs $E$ curve of $\tau_p = 0$ ps with those of $\tau_p =$ and 5 ps. It is important to note that the effect of hot phonons is two-fold. It advances the saturation of $v_d$ to occur at relatively low field $\sim 10^2$ V/cm. In addition, the hot phonon effect significantly degrades $v_{ds}$. For example, at $E = 1.5$ KV/cm, for $\tau_p = 0$, 1 and 5 ps the $v_{ds}$ are found to be 5.0, 1.6 and 1.2 $\times 10^7$ cm/s, respectively. The reduction in $v_{ds}$ is by a factor of 3.1 (4.2) for $\tau_p = 1$ (5) ps. The degradation of $v_d$ and $v_{ds}$ may be attributed to the fact that the hot phonon population increases with increasing $T_e$ [26] or $E$ which increases scattering by phonons and results into reduction in $v_d$ and $v_{ds}$. It is to be noted that the degradation of $v_{ds}$ is also clearly seen in graphene [37].

With a view to see the effect of electron density on saturation velocity, the $v_d$ vs $E$ curves are presented in figure 3 for $n_e = n_0$, $3n_0$ and $5n_0$ taking $\tau_p = 1$ ps. It is found that $v_d$ is smaller for larger $n_e$. However, the difference in magnitude is small.

We have examined the saturation velocity dependence on $n_e$, taken in range 1-10 $n_0$. The $v_{ds}$ values are taken at $E = 1.5$ KV/cm for $T = 4.2$ K and $\tau_p = 1$ ps. Expressing $v_{ds} \propto n_e^p$, we obtain $p = -0.2$. The plot of $v_{ds}$ as a function of $n_e$ is shown in figure 4 for $\tau_p = 1$ ps. It indicates that $v_{ds}$ has a weak dependence on $n_e$. Similar observation was made in graphene [32,34,36]. Moreover, we have also found that the effect of hot phonons, in degrading $v_{ds}$, is smaller for larger $n_e$ (see figure 5). For example, at $E = 1.5$ KV/cm, for $\tau_p = 1$ ps, $v_{ds}$ reduces by a factor of about 3.1 and 2.7, respectively, for $n_e = n_0$ and $5n_0$.

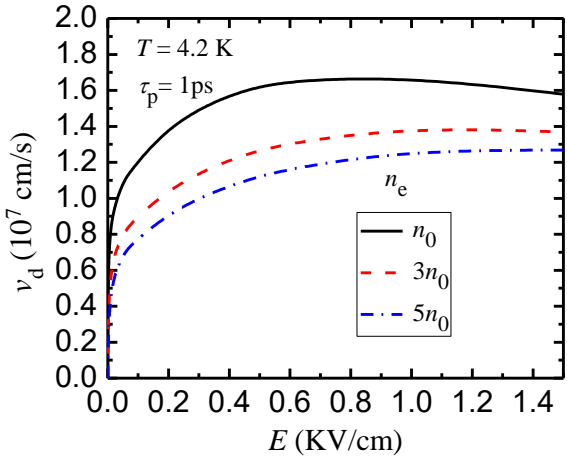

**Figure 3**. The electron drift velocity $v_d$ as a function of electric field $E$, at $T = 4.2$ K, for electron densities $n_e = 1, 3$ and $5n_0$ and phonon relaxation time $\tau_p = 1$ ps.

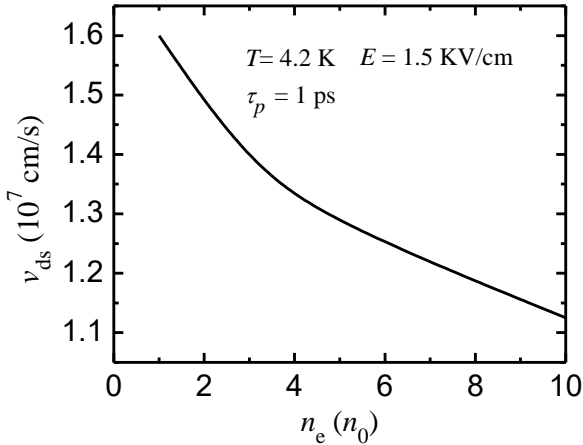

**Figure 4**. The electron saturation drift velocity $v_{ds}$ as a function of electron density $n_e$, at $T = 4.2$ K and electric field $E = 1.5$ KV/cm for phonon relaxation time $\tau_p = 1$ ps.

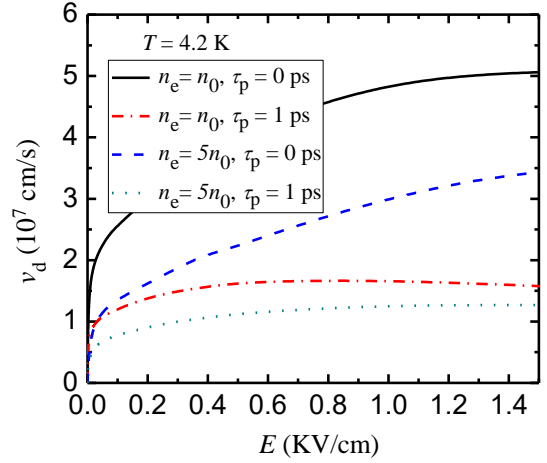

**Figure 5.** The electron drift velocity $v_d$ as a function of electric field $E$, at $T = 4.2$ K, for electron density $n_e = 1$ and $5n_0$ and phonon relaxation time $\tau_p = 0$ and 1 ps.

In order to see the effect of lattice temperature on the drift velocity, $v_d$ vs $E$ is depicted in figure 6 for three lattice temperatures $T = 4.2, 77$ and $150$ K, with $n_e = n_0$ and $\tau_p = 1$ ps. All the curves exhibit the same behavior. For a chosen $E$, the $v_d$ values are smaller for larger $T$. It may be attributed to the fact that at larger $T$, more phonons are excited causing more scattering and reducing the $v_d$. At $E = 1.5$ KV/cm, the values of $v_{ds}$ are about 1.6, 1.54 and 1.38×10$^7$ cm/s, respectively, for $T$

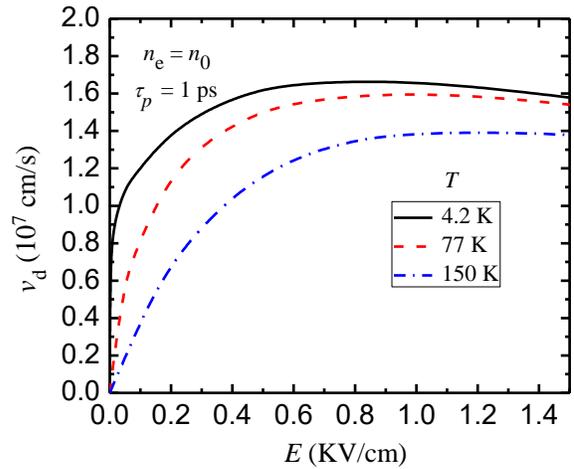

**Figure 6.** The electron drift velocity $v_d$ as a function of electric field $E$, at $T = 4.2, 77$ and $150$ K for electron density $n_e = n_0$ and phonon relaxation time $\tau_p = 1$ ps.

= 4.2, 77 and 150 K. Besides, it is noticed that the saturation of $v_d$ sets in at a higher field for larger $T$.

In the following we have discussed the current density dependence on $E$ and $n_e$. Since $j$ and $v_d$ are related by $j = n_e e v_d$, the behavior of $j$ as a function of $E$ is the same as that of $v_d$ vs $E$. However, $n_e$ dependence of $j$ is up by power 1. The saturation of current density has also been found to occur at low fields of the order $10^2$ V/cm. We have estimated saturation current density $j_s$ using $j_s = n_e e v_{ds}$. For $n_e = 1 \times 10^{18}$ cm$^{-3}$ and $\tau_p = 1$ ps, the $v_{ds}$ is about $1.6 \times 10^7$ cm/s at $E = 1.5$ KV/cm. Then, at this value of $E$, we get a large saturation current density $j_s = 2.56 \times 10^6$ A/cm$^2$. This is about 3 orders of magnitude greater than that found in conventional 3D semiconductors [29]. The $n_e$ dependence of $j_s$ can be expressed as $j_s \sim n_e^{p+1}$. With $p$ nearly equal to $-0.2$, $j_s$ increases almost linearly with increasing $n_e$. In graphene also $j_s$ has exhibited the same electron density dependence [34]. The large value of $j_s$ in 3DDS Cd$_3$As$_2$ is because of the large electron density, which is about 2-3 orders of magnitude greater than that in the conventional 3D semiconductors [29]. In graphene $j_s$ is few tens of A/cm [36,37].

The degradation of $v_{ds}$, due to the hot phonon effect, can be inhibited by reducing the phonon relaxation time $\tau_p$. In graphene, it has been shown that the hot phonon effect can be diminished by 'isotopic disorder engineering' and there by $v_{ds}$ can be enhanced [37]. This has been achieved by introducing isotropic disorder in the sample. We point out that, while disorder scattering determines the low field mobility at low temperature [21, 22], the saturation of $v_d$ has been attributed to the scattering by optical phonons. Consequently, the saturation of velocity and hence the current density are not affected by doping, In graphene, it has been shown that larger the acoustic deformation potential coupling constant $D$, larger is the $v_d$ and $v_{ds}$ [32]. This is because, scattering by acoustic phonons is dominant up to ~ 200 K as the optical phonon scattering becomes important above this temperature due to the large optical phonon energy (~ 190 meV). In Dirac semimetal Cd$_3$As$_2$ a variation of $D$ may affect $v_{ds}$ to a small extent as the optical phonon scattering becomes dominant for temperatures > ~ 40-50 K [22] and the saturation of drift velocity occurs at relatively higher temperature.

We would like to remark that, in the present investigation, el-op interaction via Fröhlich coupling with one optical phonon branch of energy 25 meV is considered although there can be numerous optical branches. A good discussion of this choice is given in our earlier work [22]. This is also evinced in the phonon mediated hot electron cooling of photoexcited carriers via pump-probe measurements [38].

## 4. Summary

Theoretically, the drift velocity $v_d$ dependence on electric field $E$ is investigated in an intrinsic three-dimensional Dirac semimetal Cd$_3$As$_2$. The hot electron mobility and energy balance equations are obtained, considering the electron scattering by acoustic and optical phonons with the screened interactions. The saturation velocity $v_{ds} \sim 10^7$ cm/s has been found at relatively small electric field (~ $10^2$ V/cm). The effects of hot phonons and electron density $n_e$ on $v_d$ and $v_{ds}$ are explored. The hot phonon effect has a strong impact on $v_{ds}$. It sets in the saturation velocity at low electric field and significantly degrades $v_{ds}$. Furthermore, $v_{ds}$ has a weak dependence on electron density. A saturation current density $j_s \sim 10^6$ A/cm$^2$ is predicted. This large $j_s$ is attributed to the large $v_{ds}$ and $n_e$ in this material. In addition, in the process of developing hot electron mobility, we have obtained the power laws for $T_e$ and $n_e$ dependence of $\mu$ in the Bloch-Grüneisen and equipartition regimes. Our theoretical predictions may be tested as and when the experimental results are available.